\documentclass[sigconf, authorversion]{acmart}

\usepackage{balance}
\usepackage{soul}
\usepackage{xcolor}
\usepackage{tabularx}
\usepackage{subcaption}
\usepackage{enumitem}
\usepackage{graphicx}
\graphicspath{{figs/}}

\newcommand{\blinduni}[1]{#1}

\AtBeginDocument{%
  \providecommand\BibTeX{{%
    \normalfont B\kern-0.5em{\scshape i\kern-0.25em b}\kern-0.8em\TeX}}}

\copyrightyear{2020} 
\acmYear{2020} 
\setcopyright{acmlicensed}
\acmConference[ACMSE 2020]{2020 ACM Southeast Conference}{April 2--4, 2020}{Tampa, FL, USA}
\acmBooktitle{2020 ACM Southeast Conference (ACMSE 2020), April 2--4, 2020, Tampa, FL, USA}
\acmPrice{15.00}
\acmDOI{10.1145/3374135.3385301}
\acmISBN{978-1-4503-7105-6/20/03}

\begin{document}

\title{Cry Wolf: Toward an Experimentation Platform and Dataset for Human Factors in Cyber Security Analysis}

\author{William Roden}
\email{roden011@gmail.com}
\affiliation{%
  \institution{University of North Carolina, Wilmington}
  \streetaddress{601 S. College Rd.}
  \city{Wilmington}
  \state{North Carolina, USA}
  \postcode{28403}
}

\author{Lucas Layman}
\email{laymanl@uncw.edu}
\orcid{0000-0002-2534-8762}
\affiliation{%
  \institution{University of North Carolina, Wilmington}
  \streetaddress{601 S. College Rd.}
  \city{Wilmington}
  \state{North Carolina, USA}
  \postcode{28403}
}


\begin{abstract}
	Computer network defense is a partnership between automated systems and human cyber security analysts. The system behaviors, for example raising a high proportion of false alarms, likely impact cyber analyst performance. Experimentation in the analyst-system domain is challenging due to lack of access to security experts, the usability of attack datasets, and the training required to use security analysis tools. This paper describes Cry Wolf, an open source web application for user studies of cyber security analysis tasks. This paper also provides an open-access dataset of 73 true and false Intrusion Detection System (IDS) alarms derived from real-world examples of \textit{impossible travel} scenarios.  Cry Wolf and the impossible travel dataset were used in an experiment on the impact of IDS false alarm rate on analysts' abilities to correctly classify IDS alerts as true or false alarms. Results from that experiment are used to evaluate the quality of the dataset using difficulty and discrimination index measures drawn from classical test theory. Many alerts in the dataset provide good discrimination for participants' overall task performance.
  
\end{abstract}



\maketitle

\section{Introduction and Background}
Computer network defense is a partnership between humans and machines. Machines are necessary to process voluminous network data, and humans must investigate suspicious behaviors identified on the network. Intrusion Detection Systems (IDSes) inspect network activity for known attack signatures, violations of user-configured rules, and statistical anomalies~\cite{Denning1987b}. IDSes then alert human cyber analysts of suspicious activity. However, IDS systems often produce false alarm rates above 95\%, requiring cyber analysts to evaluate hundreds or thousands of false alarms~\cite{Julisch2003}. This behavior may lead to a vulnerabilities at the analyst-system interface; research shows that low true event probability and high false alarm rate reduce human operator performance~\cite{Bliss2000, Breznitz1984}. While researchers strive to improve the accuracy of IDSes, comparatively little research has been published on how the \textit{behaviors} of IDSes and other semi-automated cyber security systems impact the performance of human cyber analysts.

Experimentation on cyber security analyst performance is challenging. Cyber security professionals are difficult to access due to the sensitivity of their work. Further, security-focused datasets that may be used to simulate real attacks in an experimental environment (e.g., \cite{CanadianInstituteforCybersecurity2019}) are often raw network captures and system logs rather than cyber defense tool output, and training study participants on real IDS and network monitoring systems is intractable. Consequently, few quantitative experiments have been published on analyst performance in cyber security tasks (e.g., \cite{dutt2012modeling, Ben-Asher2015}).

This paper presents two open-access resources that help address the need for controlled experimentation of cyber analyst performance. Section~\ref{sec:dataset} describes an IDS alarm dataset derived from real true and false alarms from the \blinduni{University of North Carolina, Wilmington (UNCW)}'s IDSes. Section~\ref{sec:crywolf} introduces the open source Cry Wolf web application wherein users evaluate alarms from the dataset, answer a questionnaire on user expertise, and reflect on the task. Finally, Section~\ref{sec:evaluation} presents an initial evaluation of the dataset's quality using results from a controlled experiment conducted using the Cry Wolf platform.

\section{The Impossible Travel Dataset}
\label{sec:dataset}

The first resource is a dataset of simulated IDS alerts derived from real \textit{impossible travel} alarms. Impossible travel alerts are triggered when a user authenticates from two geographic locations within a period where physical travel between the two locations is impossible, e.g., authentications from London and Moscow with a time between authentications of 30 minutes. Physical travel in this time frame is impossible, but the authentications may be legitimate through technical means such as a Virtual Private Network (VPN). 

The dataset contains \textit{true alarms} where the impossible travel alert warrants further investigation for potential malicious activity, and \textit{false alarms} where the alert is not cause for concern. The dataset contains 73 alerts in all: 30 true alarms and 43 false alarms and is available at \url{https://uncw-hfcs.github.io/ids-simulator-analysis/}. These particular numbers of alerts were needed for the experiment introduced in Section~\ref{sec:evaluation} and fully reported in \cite{Roden2019}. Each alert contains the following data:

\begin{itemize}
    \item \textbf{Cities of Authentication} --- The two geographic locations from which the IDS detected an authentication. 
    \item \textbf{Number of Successful Logins} --- The number of successful authentications from each location in the past 24 hours.
    \item \textbf{Number of Failed Logins} --- The number of failed logins from each location in the past 24 hours.
    \item \textbf{Source Provider} --- The type of internet provider the authorizations came from at each location. Possible values are:
    \begin{itemize}
        \item \textbf{Telecom} --- traditional Internet Service Providers 
        \item \textbf{Mobile/cellular} --- wireless carriers
        \item \textbf{Hosting/server} --- hosted service providers, e.g., VPNs, web hosts, and cloud computing
    \end{itemize}
    \item \textbf{Time between Authentications} --- The shortest time between authentications from the two cities in the past 24 hours.  Reported in decimal hours. This the field that triggers an alarm in a real IDS.
    \item \textbf{VPN Confidence} --- A percent likelihood that the user utilized a VPN.
\end{itemize}

Examples are provided in \S2.2. These data fields were chosen based on the first author's experience as a security operations analyst at \blinduni{UNCW}. These fields were most often considered when deciding whether an impossible travel alert was normal from the university IDSes. The data values for events were based on real samples from \blinduni{UNCW}'s IDSes with random noise added. Location pairs were generated from combinations of 12 notable cities.

\subsection{Evaluating Alerts -- the Security Playbook}
A "Security Playbook" accompanies the dataset and guides how to use the data fields to evaluate whether an alert is a true alarm or false alarm. The guidance reflects how an expert familiar with IDS behavior and the network being protected would evaluate the alerts. The playbook may be viewed at \url{https://git.io/JvE0I}. This section summarizes the main elements of the playbook as articulated to the reader.

One of the cities in each alert corresponds to legitimate access. Every city has a \textit{concern level} based on a history of attacks: Moscow and Beijing are "High" concern cities, North American cities are "Low" concern, and all others are "Medium" concern. However, users travel legitimately and hosted services may connect from other countries, thus location should not be the sole deciding factor.

\textit{Time between authentications} is the field that triggered the alert. Authentications from different locations in a short time could be an indication of account compromise. The Security Playbook provides a table of typical travel times between all pairs of locations. The \textit{ratio of successful logins to failed logins} from each location may also indicate malicious activity. More failed logins than successful logins could be an indication of password guessing, but legitimate users may fail to authenticate as well, e.g., by using an expired password. 

The \textit{source providers} help interpret the other information. A telecom provider implies the login is originating from a user physically at the location, whereas a mobile/cellular or hosting/server provider does not necessarily mean the user is physically present. There is nothing inherently safe or malicious about any type of source provider. The \textit{VPN confidence} percentage is the likelihood that the authentication attempts were made via a VPN.

The Security Playbook contains a "Things to Keep in Mind" section with additional considerations. This section reminds evaluators that IDSes can be inaccurate and that a few, many, or all of the alerts may be false alarms. Another consideration is that users visiting countries with restrictive governments will often use a VPN to circumvent that nation's firewall.
Finally, the section states that it is not unusual for a mobile device to ping the country in which the mobile device is registered when the user is traveling abroad.

\subsection{Scenarios}
The impossible travel dataset covers seven scenarios encountered in real alerts from \blinduni{UNCW}'s IDSes.

\subsubsection{True Impossible Travel}
The dataset contains 19 impossible travel alerts that are true alarms. An example is shown in Table~\ref{tab:impossible}. In these alerts, the \textit{time between authentications} is less than the time required for a person to travel between the locations as provided in the Security Playbook, and the \textit{source providers} are set to \textit{telecom} to indicate that the persons attempting the authentications are in those physical locations.

\begin{table}[ht!]
\centering
\caption{Event \#66 --- A True Alarm with Password Guessing}
\label{tab:impossible}
\begin{tabular}{@{}lrr@{}}
\toprule
City of Authentication   & Seattle     & Moscow     \\ 
\# Successful Logins     & 4           & 11         \\
\# Failed Logins         & 1           & 3          \\
Source Provider          & Telecom     & Telecom    \\ \midrule
Time between Authentications & \multicolumn{2}{c}{1.75} \\
VPN Confidence           & \multicolumn{2}{c}{0\%}  \\ \bottomrule
\end{tabular}
\end{table}

\subsubsection{Password Guessing}
The dataset contains six true alarms where the number of \textit{failed logins} from one or both cities exceeds the number of \textit{successful logins} from that city. In contrast, the ratio of failed-to-successful logins is less than 1.0 in the rest of the dataset.

\subsubsection{Edge Case Travel}
The dataset contains 15 false alarms where the \textit{time between authentications} is within 20 minutes of the typical time between the two cities. The Security Playbook emphasizes the travel time table shows "typical times" and not minima or maxima. All of the cities are "low concern" and the \textit{source providers} are either "telecom" or "mobile/cellular" to encourage focus on the times.

\subsubsection{Eurotrip}
The dataset contains six false alarms with the same features as the "edge case travel" scenarios except that both cities are in Europe. All European cities are of "medium concern", which may lead some evaluators to escalate these scenarios.

\subsubsection{Mobile}
The dataset contains eight false alarms of authentication from a mobile device. Table~\ref{tab:mobile} shows an example. A mobile device may initially route through its home country when traveling abroad as mentioned in the Security Playbook. For all the mobile scenario alerts, the \textit{source provider} is "mobile/cellular" from a city in the USA, and the other city's \textit{source provider} is "telecom". The idea is the user authenticates from a computer abroad while their mobile device is occasionally pinging their home wireless service.

\begin{table}[htb!]
\centering
\caption{Event \#18 --- A False Alarm from Mobile Usage}
\label{tab:mobile}
\begin{tabular}{@{}lrr@{}}
\toprule
City of Authentication   & Miami           & London  \\ 
\# Successful Logins     & 3               & 12      \\
\# Failed Logins         & 2               & 0       \\
Source Provider          & Mobile/Cellular & Telecom \\ \midrule
Time between Authentications & \multicolumn{2}{c}{0.90}  \\
VPN Confidence           & \multicolumn{2}{c}{0\%}   \\ \bottomrule
\end{tabular}
\end{table}

\subsubsection{VPN}
The dataset contains seven false alarms stemming from users employing VPN services in a location separate from the user's physical location. The \textit{VPN confidence} for these alerts is >90\%. One of the cities is "high concern" and has a \textit{telecom} source provider to indicate the user is physically present The other city is low or medium concern and has a \textit{hosting/service} source provider that is intended to be the VPN service.

\subsubsection{Hosting/servers}
The dataset contains 12 false alarms that entail using a server or hosting service. These alerts involve only low or medium concern cities, and one of the \textit{source providers} is "hosting/server" while the other is "telecom" to indicate the user is geographical situated in one place. The \textit{VPN confidence} values are between 53-71\% to distinguish these alarms from the higher confidence VPN scenario with high and medium concern cities.

\section{The Cry Wolf Platform}
\label{sec:crywolf}

The Cry Wolf web application provides a simulated environment for evaluating IDS alarms from the impossible travel dataset. Cry Wolf is written in the Flask micro-framework for Python. Source code and screenshots of Cry Wolf are available at \url{https://uncw-hfcs.github.io/ids-simulator/} The webapp implements an experiment to evaluate the impact of IDS \textit{false alarm rate (FAR)} on analyst performance in correctly classifying alerts as true or false alarms. The webapp provides the following experimentation structure:

\begin{enumerate}
    \item A login page with experiment description and Institutional Review Board information. User logins are assigned by a human proctor, and participants is placed into a 50\% or 86\% FAR treatment group based on the login name.
    \item A questionnaire that captures participant expertise in cyber security and networking for use in performance analysis.
    \item A training page introducing the experiment's scenario, the Security Playbook, and five training alerts with rationale.
    \item The main alert evaluation task, which displays a table of alerts and links to the alert details. The details are presented similarly to the examples shown in Tables~\ref{tab:impossible}--\ref{tab:mobile}.
    \item A post-survey that administers the NASA Task Load Index questionnaire~\cite{Hart1988} and self-reflection questions on the IDS alert evaluation for use in analysis.
\end{enumerate}

In part (4), the participant reviews the alert data against the Security Playbook and chooses to "escalate" or "don't escalate" the alert, which classifies the alert as a true or false alarm respectively. An analysis script compares participants' answers against an oracle of whether the alerts were true or false alarms, and generates a confusion matrix to derive classification performance measures of sensitivity, specificity, and precision. The repository of analysis scripts is \url{https://github.com/uncw-hfcs/ids-simulator-analysis}.

The Cry Wolf platform has several benefits for user experimentation. Nearly all of the experimental procedure is automated in the application; the only intervention required by the experimenter is to obtain informed consent and assign a pre-generated login code. The platform captures precise data on when the participants view and classify each alert, enabling fine-grained analysis of time-on-task. Subjects can exit and resume the experiment as necessary using their assigned logins, and the web application can be deployed on a cloud server for maximum availability. Cry Wolf is open source, and experimenters knowledgeable of Python and web development can adapt it to variations of the experimental structure. Further details the Cry Wolf platform structure and implementation are provided in \cite{Roden2019}.

\section{Evaluation of the Impossible Travel Dataset}
\label{sec:evaluation}

This section presents an initial evaluation of the quality of the impossible travel dataset for use in controlled experimentation of cyber analyst performance. In Fall 2019, 51 individuals participated in an experiment using the Cry Wolf platform. The goal of that experiment was to evaluate the impact of IDS \textit{false alarm rate} on analysts' abilities to \textit{correctly classify IDS alerts as true or false alarms}. The participants were randomly assigned to one of two treatment groups: 25 participants were treated with a 50\% false alarm rate (25 true and 25 false alarms), and 26 participants were treated with an 86\% false alarm rate (seven true and 43 false alarms). Participants were shown alerts from the dataset and classified each as a true or false alarm. The experiment was not time-limited, and the median time to complete the classification of all 50 alerts was 15.6 minutes. The experiment, its initial findings, and threats to validity are fully reported in~\cite{Roden2019}. The remainder of this section examines the quality of the impossible travel dataset using data from that experiment.

\subsection{Measures of Dataset Quality}
Classical test theory provides two measures to evaluate the quality of individual alerts: the alert's \textit{difficulty index} and \textit{discrimination index}~\cite{Crocker1986}. These measures are traditionally used to evaluate the quality questions in psychometric and educational tests. 

An alert's \textit{difficulty index}, $p$ is the \textit{proportion} of participants who correctly classified an alert as a true or false alarm to the total number of people who classified the alert. Lord suggests a target of $p \approx 0.85$ for items with a binary response to maximize validity of the overall test while accounting for random guessing~\cite{Lord1952}. An alert with $p=1.0$ indicates that the correct classification may be obvious to the participants, whereas an alert with a $p \ll 0.85$ may indicate that the alert data is insufficient or misleading.

An item's \textit{discrimination index}, $D$, measures the difference in responses to the item between high- and low-performers on the overall task. Each participant's number of correctly classified alerts is counted.  Two groups are formed from the 27\% of participants with the highest and lowest scores to minimize chance error~\cite{Kelley1939}. To calculate an alert's $D$, subtract the number correct in the low group from the number correct in high group, then divide the difference by the size of the larger group. Values of $D$ are in the range $[-1.0,1.0]$. Ebel states that items where $D > 0.40$ are useful discriminators, $0.20 \leq D \leq 0.40$ are in need of improvement, and $D < 0.20$ are poor discriminators and should be eliminated or rewritten~\cite{Ebel1979}.

\subsection{Analysis}

Table~\ref{tab:descriptive} shows the item difficulty and discrimination index statistics calculated per treatment group. Item difficulty ($p$) scores are similar across groups, which suggests the dataset is reliable across treatments. The mean and median discrimination index scores ($D$) are in the range warranting improvement per Ebel~\cite{Ebel1979}. The negative min $D$ value shows that at least one alert is misleading because low performers answered it correctly more than high performers.

\begin{table}[tb]
    \caption{Item Difficulty and Discrimination Index Statistics per False Alarm Rate (FAR) Treatment}
    \label{tab:descriptive}
    \centering
    \begin{tabular}{lrr|rr}
    \toprule
    {} & \multicolumn{2}{c}{50\% FAR} & \multicolumn{2}{c}{86\% FAR} \\ 
    {} &   $p$ &   $D$ &   $p$ &   $D$ \\
    mean  &       0.75 &       0.30 &       0.76 &       0.41 \\
    std   &       0.22 &       0.30 &       0.20 &       0.35 \\
    min   &       0.29 &      -0.29 &       0.16 &      -0.29 \\
    $Q_1$   &       0.60 &       0.00 &       0.62 &       0.14 \\
    $Q_2$ (median)   &       0.76 &       0.29 &       0.77 &       0.43 \\
    $Q_3$   &       0.95 &       0.57 &       0.92 &       0.71 \\
    max   &       1.00 &       0.86 &       1.00 &       1.00 \\
    \midrule
    {participants} & \multicolumn{2}{c|}{25} & \multicolumn{2}{c}{26} \\
    {true alarms} & \multicolumn{2}{c|}{25} & \multicolumn{2}{c}{6} \\
    {false alarms} & \multicolumn{2}{c|}{25} & \multicolumn{2}{c}{44} \\ 
    \bottomrule
    \end{tabular}
    
\end{table}

Table~\ref{tab:assessment} bins the alerts according to their $p$ and $D$ scores. Approximately 40\% of the alerts evaluated by each group have a good discrimination index per Ebel~\cite{Ebel1979}. Approximately 30\% of the alerts fall into the "too easy" category where many of the participants correctly classified the alert. In the Cry Wolf experiment, unlike in educational tests, such alerts are useful because they help build and reinforce the participants' notions of true and false alarms.

\begin{table}[b]
    \caption{Aggregate Measures of Item Analysis. Values are the Number of Alerts.}
    \label{tab:assessment}
    \centering
    \begin{tabular}{lrr}
    \toprule
    {} &  50\% FAR &  86\% FAR \\
    \midrule
$D > 0.4$ (best)                         &        24 &        27 \\
$p \geq Q_3$ and $D \leq 0.4$ (too easy) &        13 &        17 \\
$p < Q_2$ and $D \leq 0.4$ (too hard)    &         7 &         2 \\
    \bottomrule
    \end{tabular}
\end{table}

The alarms that are "too hard" are cause for concern as they may indicate that the Cry Wolf training exercises are insufficient or the alerts are unclear. Two \textit{eurotrip} alerts have travel times that are nearly impossible but are false alarms -- likely the participants were erring on the side of caution. Four alerts from the \textit{mobile} scenario fell into the "too hard" category for the 50\% FAR group. One possible explanation is that the participants did not read or remember the line near the end of the Security Playbook stating that mobile devices abroad often ping their home city first. The other alerts in the "too hard" group exhibit no obvious pattern.  The $D$ score is sensitive to random noise given the small sizes of the high and low groups ($n = 7$ in each group) for each treatment.

Overall, the difficulty and discrimination indices suggest that the impossible travel dataset is useful to discriminate between high- and low- performers.  Further qualitative investigation on why the "too hard" alerts were mis-classified is warranted.  

\section{Conclusion and Future Work}

This paper presents a dataset of simulated IDS alarms, introduces the Cry Wolf platform for controlled experiments of cyber analyst performance, and provides an initial evaluation of the dataset's quality. The impossible travel dataset and Cry Wolf platform are open source, and the evaluation shows promising results for discriminating between high and low performers. 

In future work, the impossible travel dataset will be expanded to include new alarm types. The Cry Wolf platform will be used in Summer 2020 to study the effects of \textit{anchoring bias} \cite{Tversky1974} on cyber analyst performance. These studies are part of a research plan to investigate factors impacting cyber analyst performance. The long term goals of this research are to develop new training methodologies for coping with imperfect cyber defense systems, provide guidance for tool developers on user interface considerations, and develop defense strategies against analyst-machine vulnerabilities.

\balance
\bibliographystyle{ACM-Reference-Format2}
\bibliography{references, thesis}

\appendix

\end{document}